\documentclass[reprint,aps,prb,english,nolongbibliography,superscriptaddress]{revtex4-2}

\usepackage{graphicx,dcolumn}
\usepackage[table,xcdraw,dvipsnames]{xcolor}
\usepackage[colorlinks=true,citecolor=blue,urlcolor=blue,linkcolor=blue]{hyperref}

\usepackage{amsmath,amssymb,bm,mathrsfs,amsfonts}

\newcommand{\pd}[2]{\frac{\partial #1}{\partial #2}}

\newcommand{\bra}[1]{\langle #1 |}
\newcommand{\ket}[1]{| #1 \rangle}

\newcommand{\av}[1]{\langle #1 \rangle}

\let\tilde\relax
\newcommand{\tilde}[1]{\widetilde{#1}}
\let\Re\relax

\DeclareMathOperator{\Re}{Re}

\newcommand{\bS}{\boldsymbol{S}}

\newcommand{\mH}{\mathcal{H}}
\newcommand{\mC}{\mathcal{C}}
\newcommand{\mA}{\mathcal{A}}

\newcommand{\eqn}[1]{\begin{equation} #1 \end{equation}}

\begin{document}

\title{Subdiffusive {spin} transport in disordered classical Heisenberg chains}

\author{Adam J. McRoberts}
\email{amcr@pks.mpg.de}
\affiliation{Max Planck Institute for the Physics of Complex Systems, N\"othnitzer Str.\ 38, 01187 Dresden, Germany}

\author{Federico Balducci}
\email{federico.balducci@uni.lu}
\affiliation{Department of Physics and Materials Science, University of Luxembourg, L-1511 Luxembourg, Luxembourg}
\affiliation{SISSA, via Bonomea 265, 34136, Trieste, Italy}

\author{Roderich Moessner}
\affiliation{Max Planck Institute for the Physics of Complex Systems, N\"othnitzer Str.\ 38, 01187 Dresden, Germany}

\author{Antonello Scardicchio}
\affiliation{ICTP, Strada Costiera 11, 34151, Trieste, Italy}
\affiliation{INFN Sezione di Trieste, Via Valerio 2, 34127 Trieste, Italy}

\date{\today}

%--------------------------------------------------------------------------------------------------
%--------------------------------------------------------------------------------------------------

\begin{abstract}
	We study the transport and equilibration properties of a classical Heisenberg chain, whose couplings are random variables drawn from a one-parameter family of power-law distributions. The absence of a scale in the couplings makes the system deviate substantially from the usual paradigm of diffusive spin hydrodynamics, and exhibit a regime of subdiffusive transport with an exponent changing continuously with the parameter of the distribution. We propose a solvable phenomenological model that correctly yields the subdiffusive exponent, thereby linking local fluctuations in the coupling strengths to the long-time, large-distance behaviour. It also yields the finite-time corrections to the asymptotic scaling, which can be important in fitting the numerical data. We show how such exponents undergo transitions as the distribution of the coupling gets wider, marking the passage from diffusion to a regime of slow diffusion, and finally to subdiffusion.  
\end{abstract}

\maketitle

%--------------------------------------------------------------------------------------------------
%--------------------------------------------------------------------------------------------------
\section{Introduction}

Ever since the original recognition that diffusion is absent in certain random lattices~\cite{Anderson1958Absence}, the study of transport in impure materials has been a rich source of surprising, and often subtle, phenomena. Recently, the study of the mechanism of equilibration in quantum many-body systems has provided an additional impetus, carried by experimental advances~\cite{Leibfried2003Quantum,Bloch2008Many,Ueda2020Quantum} as well as concomitant conceptual progress~\cite{Deutsch1991Quantum,Srednicki1994Chaos,Basko2006Metal}. The quantum statistical mechanics of non-equilibrium systems, and of the process of equilibration itself~\cite{DAlessio2016Quantum,Abanin2019Colloquium}, is now reaching the level of detail that classical ergodic theory has reached more than a hundred years after the works of Boltzmann~\cite{Cornfeld1982Ergodic,Vulpiani2009Chaos}.

It is, in particular, the study of systems with both interactions and disorder that has thrown up many puzzles. This is subject to formidable technical difficulties, as exact solutions are generically unavailable, whilst numerics for quantum systems is typically restricted to small system sizes and/or short times. This has led to vigorous debates regarding nature and lifetime of possible intermediate-time dynamical regimes~\cite{Suntajs2020Quantum,Sierant2020Thouless,Panda2020Can,Abanin2021Distinguishing,Morningstar2022Avalanches,Crowley2022Constructive,Sierant2022Challenges} (and the role of rare events in their genesis~\cite{Agarwal2015Anomalous,DeTomasi2021Rare}), and how to distinguish them from expected or desired long-time behaviour. 

One aspect of much recent interest relates to the question of under what conditions, and with what consequences, many-body systems may exhibit neither diffusive nor localised behaviour; much-explored possibilities relate to subdiffusive~\cite{Znidaric2016Diffusive,Schulz2018Energy,Mendoza201Asymmetry,Taylor2021Subdiffusion} or Kardar-Parisi-Zhang~\cite{Kardar1986Dynamic,Ljubotina2017Spin,Ljubotina2019Kardar,Dupont2020Universal,DeNardis2020Universality} behaviours.

Here, we study a family of disordered, classical chains of Heisenberg spins. This picks up the aforementioned threads in the following ways. (i) Such chains are, a priori, generic one-dimensional many-body systems, but (ii) even the clean (i.e. without disorder) Heisenberg chain has recently been shown to be capable of exhibiting extended non-diffusive transport regimes~\cite{Bilitewski2021Classical,McRoberts2022Anomalous,McRoberts2022Long,McRoberts2022Prethermalization}---closing a long-standing debate regarding the diffusive nature of excitations~\cite{Muller1988Anomalous,Gerling1990Time,Liu1991Deterministic,DeAlcantara1992Breakdown,Srivastava1994Spin,Constantoudis1997Nonlinear,Bagchi2013Spin,Li2019Energy}. (iii) It is technically possible to simulate large system sizes for long times, and thus there is hope of probing various regimes and their crossovers, all the more as (iv) the tuning parameter distinguishing members of the family of models allows us to access very different behavioural regimes. 

In the following, we show how this family of disordered Heisenberg chains exhibits a rich set of transport phenomena, comprising standard diffusion as well as tunable subdiffusion, but, as established in previous work~\cite{Oganesyan2009Energy,Richter2020Decay}, no classical counterpart of many-body localisation. We account for all of these phenomena with a relatively simple treatment, which makes transparent the role of extreme-value statistics and rare phenomena. We also provide detailed insights into the origin and nature of short- to intermediate-time crossovers and corrections, which can be important in the numerics over a broad time window.

Atypical {\it rare regions} of the system, where, for example, local couplings are much smaller or much larger than their typical value, are suspected to play a significant role in achieving or inhibiting thermalisation in classical and quantum systems. However, their signature in  numerical results is often obscured, and can lead to different, contrasting interpretations of finite-size and finite-time numerics on account of the very slow emergence of the true asymptotic behaviour. It is therefore highly desirable to have access to models in which analytical results sufficiently constrain the data analysis to yield a clear interpretation of the numerics. 

In this work, we present one such example. Crucially, our solution of a phenomenological model for transport, in which the local diffusion coefficient is a broadly distributed random variable, provides both leading {\it and subleading} terms in the large-time expansion of observables. We show how, in the absence of such an analytic prediction for the subleading behaviour, slow diffusion could be mistaken for subdiffusion, the diffusive term achieving dominance for times orders of magnitude larger than those typically reachable in state-of-the-art numerics.

The paper is organised as follows. In \S\ref{sec:model} we introduce the family of models we study. In \S\ref{sec:numerics} we present the numerical results, and explain how to pin down the diffusion/subdiffusion transition from them. In \S\ref{sec:effective_model} we develop an effective model for the dynamics of the Heisenberg chain, upon which the understanding of the numerical results is based. Finally, in \S\ref{sec:conclusions} we draw our conclusions. Additional information regarding the effective model is provided in Apps.~\ref{app:sec:integral_eq} \& \ref{app:sec:effective_model}.

%--------------------------------------------------------------------------------------------------
%--------------------------------------------------------------------------------------------------
\section{Model}
\label{sec:model}

\begin{figure}
    \centering
    \includegraphics[width=\columnwidth]{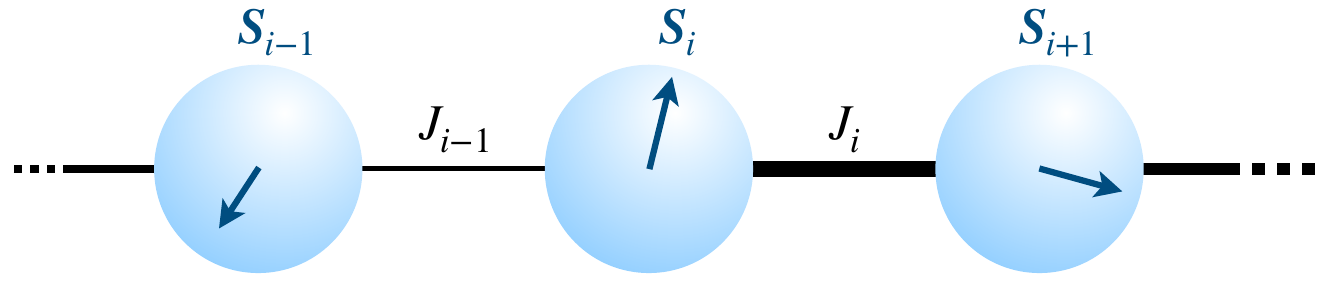}
    \caption{Sketch of the spin chain from Eq.~\eqref{eq:Heisenberg_Hamiltonian}. The spins $\bS_i$ live on the sphere $S^2$, and interact via broadly-distributed nearest-neighbour couplings $J_i$, cf. Eq.~\eqref{eq:pdf_J}. We use periodic boundary conditions.}
    \label{fig:spinchain}
\end{figure}

\begin{figure}[t]
    \centering
    \includegraphics{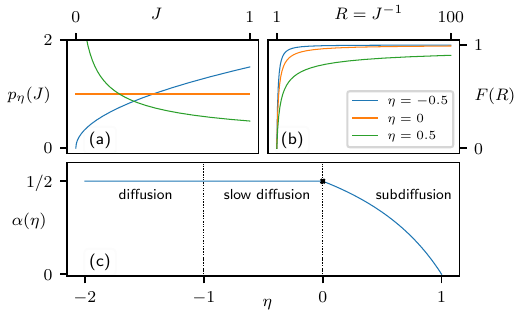}
    \caption{(a) Probability density function of the couplings $J_i$ for representative values of $\eta$. (b) Cumulative distribution function of the inverse couplings $R = J^{-1}$, showing the fat tails of the distribution for $\eta > 0$. (c) Overview of the dynamical regimes as a function of $\eta$. The point at $\eta = 0$ corresponds to logarithmically-suppressed diffusion.}
    \label{fig:overview}
\end{figure}

We consider a bond-disordered version of the classical Heisenberg chain, with the Hamiltonian
\eqn{
    \mH = \sum_{i=1}^L J_i \bS_i \cdot \bS_{i+1},
    \label{eq:Heisenberg_Hamiltonian}
}
where $\bS_i \in S^2$ are classical unit-length spins, and we use periodic boundary conditions. The random couplings $J_i$ are independent and identically distributed (i.i.d.), and drawn from a one-parameter family of power-law distributions. The probability density function,
\eqn{
    \label{eq:pdf_J}
    p_{\eta}(J) = (1 - \eta) J^{-\eta}, \qquad J \in [0, 1]
}
is controlled by an exponent $\eta \in (-\infty, 1)$. We show the distributions $p_{\eta}(J)$ for representative values of $\eta$ in Fig.~\ref{fig:overview}: for $\eta > 0$, the probability density diverges at $J = 0$; for $\eta < 0$, the weight accumulates around $J = 1$, and approaches the clean model as $\eta \rightarrow -\infty$; precisely at $\eta = 0$, the distribution is uniform. Throughout, units are implicitly defined by the maximum coupling $J_{\mathrm{max}} = 1$.

It is, however, the distribution of the \textit{inverse} couplings $R = J^{-1}$ that determines the bare dynamical timescales. Their probability density,
\eqn{
    \label{eq:pdf_R}
    q_{\eta}(R) = (1 - \eta)R^{\eta - 2},\qquad
    R \in [1, \infty), 
}
is fat-tailed: the first moment $\overline{R}$ diverges for $\eta > 0$; the second moment $\overline{R^2}$ diverges for $\eta > -1$; and so on for the higher moments (we denote the average over disorder by an overline). We will derive the consequences of these divergences in $\S${\ref{sec:effective_model}}.

Now, the classical dynamics of the Hamiltonian (\ref{eq:Heisenberg_Hamiltonian}) is defined by the fundamental Poisson brackets,
\eqn{
\{S_i^\alpha, S_j^\beta\} = \delta_{ij} \epsilon^{\alpha \beta\gamma} S_i^\gamma,
}
from which one obtains the equations of motion,
\eqn{
    \partial_t \bS_i = (J_{i-1}\bS_{i-1} + J_i \bS_{i+1}) \times \bS_i.
    \label{eq:eom}
}
These equations are manifestly $\mathrm{SO}(3)$ invariant: as in the clean model, all three components of the magnetisation are conserved.

%--------------------------------------------------------------------------------------------------
%--------------------------------------------------------------------------------------------------
\section{Numerical results for the dynamics}
\label{sec:numerics}

\begin{figure*}[t]
    \centering
    \includegraphics{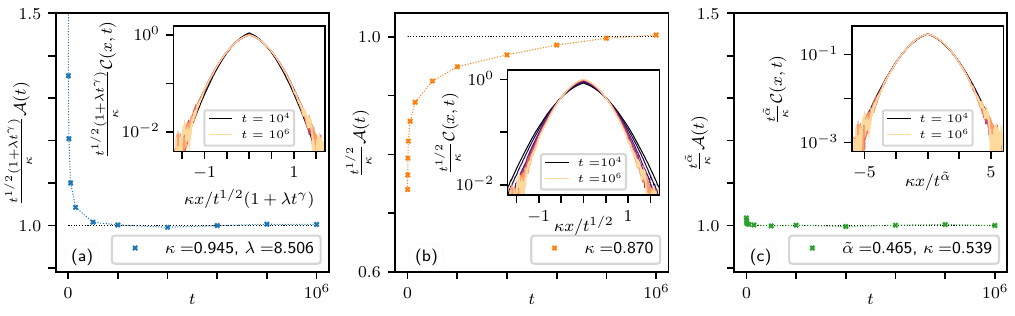}
    \caption{Spin dynamics in the ``slow diffusion'' regime, shown for $\eta = -0.5$. In each panel, the main figure shows the rescaled autocorrelator $\mA(t) = \mC(x=0,t)$ (cf. Eq.~\eqref{eq:corr}). The insets show the corresponding scaling collapse of the full correlation function $\mC(x, t)$.
    (a) Autocorrelator and scaling collapse from Eq.~\eqref{eq:slow_diffusion_autocorr}, i.e., assuming an asymptotic diffusive behaviour with strong, anomalous corrections.
    (b) Diffusive scaling without finite-time corrections (i.e., setting $\lambda = 0$, a one-parameter fit), showing that the corrections must be accounted for, at least up to the final time of the simulation, $t = 10^6$.
    (c) Numerical fit with an anomalous (subdiffusive) exponent, i.e., a direct two-parameter fit to Eq.~(\ref{eq:autocorr}). See main text for additional details.}
    \label{fig:slow_diffusion}
\end{figure*}

We study the dynamics of the model~\eqref{eq:Heisenberg_Hamiltonian} at infinite temperature; in particular, we consider the correlation function of the spins,
\eqn{
    \mC(j,t) := \overline{ \av{\bS_j(t)\cdot\bS_0(0) } } ,
\label{eq:corr}
} 
and the associated autocorrelator $\mA(t) := \mC(0, t)$, averaged over both realisations of disorder (overline) and a thermal ensemble of initial states (angular brackets), as described below. In the long-time limit, we expect to reach a hydrodynamic regime, wherein the correlation functions of conserved densities approach an asymptotic scaling form,
\eqn{
\mC(x, t) \sim t^{-\alpha} \mathcal{F}(x/t^{\alpha}),
}
for some universal function $\mathcal{F}$ and scaling exponent $\alpha$. The latter can also be obtained by fitting the autocorrelator to a power law:
\eqn{
\mA(t) \simeq \kappa t^{-\alpha}.
\label{eq:autocorr}
}
However, the asymptotics---whilst they define the dynamical exponent---capture only the leading behaviour. As we will show, the finite-time corrections to Eq.~(\ref{eq:autocorr}) can be quite severe, and persist, at least, to late times $t = 10^6$ (in units with $J_{\max}=1$). 

 To evaluate the correlator (\ref{eq:corr}) for a given disorder exponent $\eta$, we construct an ensemble of 20000 initial states at infinite temperature: each spin is simply, and independently, drawn from the uniform distribution on the sphere. For each state in the ensemble, we draw a distinct realisation of the couplings $J_i$, and numerically integrate the equations of motion (\ref{eq:eom}). Snapshots of the state are stored at intervals of $\Delta t = 10$, with the correlation function at a given time-difference $t$ calculated by averaging over 1000 consecutive snapshots. Data shown are for the system size $L = 8192$~\footnote{We have simulated system sizes $L = 2048$, $4096$, $8192$ and found no appreciable differences in the data, indicating that the numerics are principally limited by finite-time effects, not finite-size effects.}. 

There are four distinct dynamical regimes. First, in the clean limit ($\eta \rightarrow -\infty$), the spin dynamics is diffusive. Even in this limit, however, finite-size and finite-time effects are capable of hiding the asymptotic behaviour, and it has taken modern-day computing resources~\cite{Bagchi2013Spin,Li2019Energy,Bilitewski2021Classical,McRoberts2022Anomalous} to resolve the long-lasting debate on this topic~\cite{Muller1988Anomalous,Gerling1990Time,Liu1991Deterministic,DeAlcantara1992Breakdown,Srivastava1994Spin,Constantoudis1997Nonlinear}. Second, as $\eta$ becomes finite, and in particular for $-1 \leq \eta < 0$, diffusion persists at extremely large times, but finite-time corrections become increasingly severe. This is due to the existence of local dynamical bottlenecks, which arise from the growing probability of drawing an arbitrarily small coupling. We refer to this regime as ``slow diffusion'', and study it in detail in \S\ref{sec:numerics_slow_diffusion}. Third, at $\eta = 0$, the probability density $p_{\eta}(J) = 1$ becomes uniform, and the first moment of the inverse couplings, i.e., $\overline{R}$, diverges logarithmically. Accordingly, the asymptotic spin dynamics shows logarithmically-suppressed diffusion, cf. \S\ref{sec:numerics_logarithmic_suppression}. Finally, when $0 < \eta < 1$, the spin dynamics is truly subdiffusive, with an exponent (Hurst index) $\alpha < 1/2$, cf. \S\ref{sec:numerics_subdiffusion}.

In practise, for all the considered cases, the leading corrections to the asymptotics are required to obtain the correct scaling exponent $\alpha$ from the numerical data. If the corrections are neglected, one finds an $\alpha$ smaller than the true value. We first present the numerical results, and develop an effective model which accounts for our observations in $\S${\ref{sec:effective_model}}.

\begin{figure*}[t]
    \centering
    \includegraphics{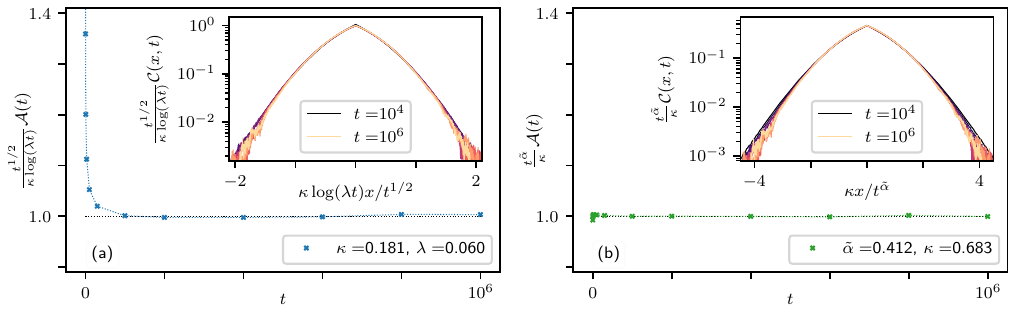}
    \caption{Spin dynamics at the slow-diffusion/subdiffusion transition, $\eta = 0$.
    (a) Rescaled autocorrelator (main panel) and scaling collapse of the correlation function (inset), as predicted by the logarithmic suppression of diffusion, i.e., a two-parameter fit to Eq.~\eqref{eq:log_subdiffusion_autocorr}.
    (b) Same data as panel (a), but fitting with a subdiffusive exponent, i.e., a direct two-parameter fit to Eq.~\eqref{eq:autocorr}. Note that the scaling collapse in the tails of the correlations is slightly better using the logarithmic-suppression, indicating that this is the correct picture.}
    \label{fig:log-subdiffusion}
\end{figure*}

%--------------------------------------------------------------------------------------------------
\subsection{\boldmath$-1 \leq \eta < 0$: Slow diffusion}
\label{sec:numerics_slow_diffusion}

We begin with the regime of slow diffusion, observed when $-1 \leq \eta < 0$. Here, $p_{\eta}(J)$ is maximal at $J = 1$ and vanishes at $J = 0$ (cf.\ Fig.~\ref{fig:overview}).

In this regime, the average bare timescale $\overline{R}$ remains finite, so the leading behaviour remains diffusive, i.e., $\alpha = 1/2$. Higher moments (e.g. $\overline{R^2}$), however, diverge, giving rise to strong corrections. Correspondingly, the autocorrelator takes the form
\eqn{
    \mA(t) \sim \frac{\kappa}{t^{1/2}(1 + \lambda t^{\gamma})}, \qquad
    \gamma = \frac{\eta}{1 - \eta} < 0,
\label{eq:slow_diffusion_autocorr}
}
with $\kappa$ and $\lambda$ obtained numerically in a two-parameter fit, and the scaling function $\mathcal{F}$ approaches a Gaussian at late times. The subleading exponent $\gamma$ is fixed by the effective model of \S\ref{sec:effective_model}. We show the slow-diffusion dynamics for a representative value $\eta = -0.5$ in Fig.~\ref{fig:slow_diffusion}. We find that the corrections postulated by Eq.~\eqref{eq:slow_diffusion_autocorr} capture the slow spreading of the correlations, and are necessary to obtain the correct scaling at finite times. 

It is interesting to point out that it is also possible to fit the correlations with an anomalous (subdiffusive) exponent, i.e., applying a two-parameter fit to Eq.~(\ref{eq:autocorr}) directly. In the slow diffusion regime, this procedure yields a numerical agreement with the simulation data that is comparable to the diffusion-with-strong-corrections hypothesis. We stress, however, that subdiffusion is \emph{not} the correct asymptotic picture; rather, it is an artefact of the corrections taking the form of a sum of (small) power-laws. Indeed, at $\eta = -0.5$, and for the times accessible by our numerics, a direct power-law fit finds a subdiffusive exponent $\alpha = 0.465$. The effective model we develop in \S\ref{sec:effective_model} instead predicts the form Eq.~\eqref{eq:slow_diffusion_autocorr} with $\gamma = -1/3$: plugging in the value of $\lambda$ found from the fits, it holds exactly that
\begin{equation}
    \pd{\log \mA(t)}{\log(t)} \approx 0.465\dots \qquad \text{at } t = 10^5.
\end{equation}
Thus, whilst locally, around the largest times we could access, the effective power-law decay is slower than $1/2$, our analytical understanding predicts that this is but a crossover, and much longer times ($t \approx 10^8$) are needed for the corrections to become negligible (say, 1\% of the leading term).

%--------------------------------------------------------------------------------------------------
\subsection{\boldmath$\eta = 0$: Logarithmically-suppressed diffusion}
\label{sec:numerics_logarithmic_suppression}

At $\eta = 0$ the slow diffusion regime terminates. The distribution of the couplings $J_i$ becomes uniform over $[0, 1]$, and the corrections to the diffusive behaviour are enhanced, fundamentally changing the leading asymptotics. In particular, spin diffusion is now logarithmically suppressed, and one finds
\eqn{
\mA(t) \sim \frac{\kappa \log(\lambda t)}{t^{1/2}},
\label{eq:log_subdiffusion_autocorr}
}
cf.\ \S\ref{sec:corrections_scaling}.

We show the logarithmically-suppressed diffusion in Fig.~\ref{fig:log-subdiffusion}. Whilst, again, a direct fit to Eq.~\eqref{eq:autocorr}---i.e., a fit to determine the subdiffusive exponent---is in good agreement with the data (Fig.~\hyperref[fig:log-subdiffusion]{\ref{fig:log-subdiffusion}(b)}), in this case the corresponding scaling collapse is slightly worse in the tails than that provided by the logarithmic-suppression picture (Fig.~\hyperref[fig:log-subdiffusion]{\ref{fig:log-subdiffusion}(a)}). That Eq.~\eqref{eq:log_subdiffusion_autocorr} fits both the centre (the autocorrelator) and the tails of the correlations is strong evidence in favour of the picture predicted by the effective model of \S\ref{sec:effective_model}. 

From Fig.~\ref{fig:log-subdiffusion} one can also appreciate how the scaling function $\mathcal{F}$ is no longer Gaussian, and has developed a non-analytic feature at the origin---which will become more pronounced in the subdiffusive regime.

%--------------------------------------------------------------------------------------------------
\subsection{\boldmath$\eta > 0$: Subdiffusion}
\label{sec:numerics_subdiffusion}

Finally, we turn to the case $\eta > 0$. The distribution $p_{\eta}(J)$ now diverges at $J = 0$, which means that a finite fraction of the bonds become arbitrarily small. This leads to truly subdiffusive dynamics, with exponent $\alpha < 1/2$. 

\begin{figure*}[t]
    \centering
    \includegraphics{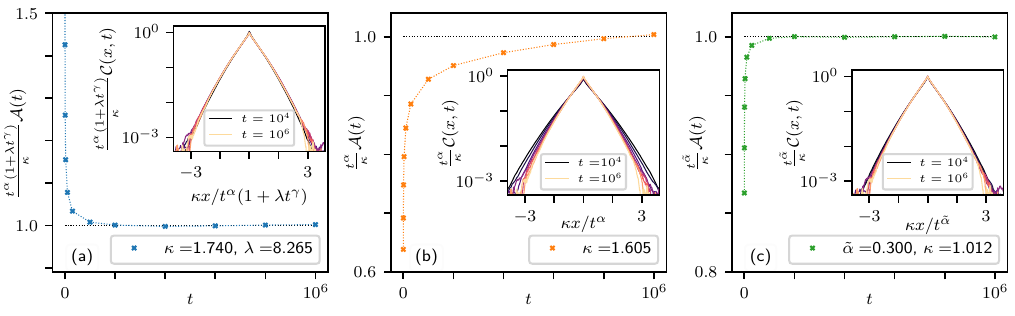}
    \caption{Subdiffusive spin dynamics in the strongly-disordered regime, shown for a representative $\eta = 0.5$. Again, each main panel shows the autocorrelator, whilst the insets show the scaling collapse of the correlation function.
    (a) The autocorrelator and scaling collapse from Eq.~(\ref{eq:subdiffusion_autocorr}), i.e., a two-parameter fit with the subdiffusive exponent $\alpha = (1 - \eta)/(2 - \eta) = 1/3$, and with leading corrections included.
    (b) Subdiffusion with the same exponent as panel (a), but without the corrections, i.e., a one-parameter fit setting $\lambda = 0$.
    (c) Scaling with an exponent obtained numerically without finite-time corrections, i.e., a direct two-parameter fit to Eq.~\eqref{eq:autocorr}. Again, note that the scaling collapse in the tails is better in (a) than in (c), indicating that the finite-time corrections provide the correct picture.}
    \label{fig:subdiffusion}
\end{figure*}

However, the fact that corrections in the slow diffusion regime were strong enough that a naive numerical fit to Eq.~(\ref{eq:autocorr}) already finds subdiffusion at $\eta < 0$ suggests that, again, there will be strong corrections which hide the correct exponent (the exponent obtained numerically is continuous as a function of $\eta$). This is indeed the case, and we find 
\begin{equation}
    \mA(t) \sim \frac{\kappa}{t^{\alpha}(1 + \lambda t^{\gamma})},
\label{eq:subdiffusion_autocorr}
\end{equation}
cf.\ \S\ref{sec:corrections_scaling}, with
\begin{equation}
    \alpha = \frac{1 - \eta}{2 - \eta}, \qquad
    \gamma = 2\alpha - 1 < 0.
\end{equation}
We show the subdiffusive dynamics for $\eta = 0.5$ in Fig.~\ref{fig:subdiffusion}, again finding that the leading finite-time corrections are required. As was the case at $\eta = 0$, we show that the the form of $\mA(t)$ predicted by the effective model, Eq.~\eqref{eq:subdiffusion_autocorr}, collapses the tails of the correlations slightly better than a direct fit to Eq.~(\ref{eq:autocorr}) (compare Figs.~\hyperref[fig:subdiffusion]{\ref{fig:subdiffusion}(a)} and \hyperref[fig:subdiffusion]{\ref{fig:subdiffusion}(c)}).

As to the scaling function $\mathcal{F}$, it is clear from Fig.~\ref{fig:subdiffusion} that it is not Gaussian. It appears from the inset scaling collapses that $\mathcal{F}$ approaches a stretched-exponential; though we cannot draw a sharp conclusion on this point since our simulations do not reach the truly asymptotic regime.

%--------------------------------------------------------------------------------------------------
%--------------------------------------------------------------------------------------------------
\section{An effective model for spin transport}
\label{sec:effective_model}

Having first presented the numerical data, including, without justification, the forms of the corrections, we now present a phenomenological model of the spin transport which explains, at least qualitatively, the results of $\S${\ref{sec:numerics}}. In particular, in \S\ref{sec:effective_model_motivation} we introduce the effective model and motivate its form. In \S\ref{sec:transfer_matrix} we show how the model can be solved by means of a transfer-matrix technique. Finally, in \S\ref{sec:leading_scaling} we extract the asymptotic scaling via the transfer-matrix trick, whilst in \S\ref{sec:corrections_scaling} we address the finite-time corrections to the asymptotics. The reader may refer to App.~\ref{app:sec:integral_eq} for another way of solving the effective model.

%--------------------------------------------------------------------------------------------------
\subsection{Motivation for the effective model}
\label{sec:effective_model_motivation}

In order to access the late-time behaviour of the spin-spin correlations, we set ourselves on a hydrodynamic scale, and linearise the microscopic dynamics whilst accounting for the local exchange of energy and spin. By rotational invariance, we consider only the magnetisation density along one axis: whilst one could write down hydrodynamic equations that couple the local magnetisation components along all three axes in an SO(3)-symmetric fashion, we will show that the simplest ansatz of totally decoupled components is sufficient to explain our numerical findings. 

Let us denote the magnetisation at the coarse-grained site $x$ by $m_x(t)$. We retain a discretised lattice even on the hydrodynamic scale, because this way it is easier to account for the strong, fat-tailed disorder of Eq.~\eqref{eq:pdf_J}. The disorder in the couplings $J_x$ suggests that the local diffusion coefficient $D_x$ will vary similarly, giving rise to a local diffusion equation: 
\begin{equation}
    \label{eq:random_diffusion}
    \partial_t m_x = D_{x-1} m_{x-1} + D_x m_{x+1} - (D_x + D_{x+1}) m_x.
\end{equation}
We argue that this is the correct lattice discretisation of the diffusion equation, since it comes from enforcing Kirchhoff's law at each site, i.e., the inflow and outflow of magnetisation at each site cancel out. In turn, this fact implies that the constant vector $m_x \equiv m$ is a stable solution of Eq.~\eqref{eq:random_diffusion}, and that the magnetisation is locally conserved.

The local, random diffusion coefficients $D_x$ have some unknown distribution function, depending on the underlying $J_i$'s. Interpreting Eq.~\eqref{eq:random_diffusion} as a coarse graining of Eq.~\eqref{eq:eom}, the distribution of the $D_x$'s should be obtained, in principle, via some renormalisation procedure (mayhaps akin to the strong-disorder renormalisation group used for quantum disordered spin chains~\cite{Ma1979Random,Dasgupta1980Low,Fisher1995Critical,Refael2013Strong,Igloi2018Strong}). For our purposes, however, it suffices to assume that $1/D_x$ has the same fat tails as $1/J_i$. We thus assume, for simplicity, that $D_x$ has the same probability density function as in Eq.~\eqref{eq:pdf_J}:
\begin{equation}
    p_\eta(D)=(1-\eta)D^{-\eta}, \qquad D \in [0,1].
\end{equation}
Using a different distribution that shares the same tails leads to equivalent results, as will become clear from the extreme-value analysis below. We have, here, implicitly rescaled the units of time by setting the maximum possible value $D_{\max} = 1$.

In the following, we solve the dynamics described by Eq.~\eqref{eq:random_diffusion}. Before doing so, however, some comments are in order. First, we stress that Eq.~\eqref{eq:random_diffusion} was already known to approximate the dynamics of the classical Heisenberg model \emph{near zero temperature}~\cite{Alexander1981Excitation,Haus1987Diffusion,Bouchaud1990Anomalous}. Indeed, a spin-wave expansion of the equations of motion~\eqref{eq:eom} leads to a copy of Eq.~\eqref{eq:random_diffusion}, if the interactions among spin waves are neglected. Our new contribution is to show that this linearised equation is also capable of describing the dynamical scaling of space and time \textit{at infinite temperature}.

Second, we remind the reader that Eq.~\eqref{eq:random_diffusion} has been the subject of a vast literature, pioneered by Dyson~\cite{Dyson1953Dynamics} (see also the reviews~\cite{Alexander1981Excitation,Haus1987Diffusion,Bouchaud1990Anomalous}), and sometimes goes by the name of ``random barrier model''. The long-time behaviour of $m_x(t)$ can be obtained by various methods: an integral equation that leads to the exact solution~\cite{Dyson1953Dynamics,Bernasconi1980Diffusion,Nieuwenhuizen1985Transport}, small-disorder expansions~\cite{Derrida1983Diffusion,Denteneer1984Diffusion}, an effective-medium theory~\cite{Alexander1981Excitation}, and renormalisation-group approaches~\cite{Machta1981Generalized,Machta1983Renormalization,Guyer1984Diffusion,Machta1985Random}. Here, we solve the model by yet another technique---a series expansion of a transfer-matrix representation---for two reasons: first, we find it faster, and more transparent from a physical standpoint; second, it allows us to access the subleading terms, which, as noted in $\S$\ref{sec:numerics}, must be taken into account. In the following sections, we describe in detail the transfer-matrix method, and then extract the scaling in the different dynamical regimes. We benchmark our solution against the integral equation method in App.~\ref{app:sec:integral_eq}.

%--------------------------------------------------------------------------------------------------
\subsection{Transfer matrix representation, and solution in the clean case}
\label{sec:transfer_matrix}

The exact solution of Eq.~\eqref{eq:random_diffusion} can be obtained only numerically, owing to the random nature of the $D_x$'s. However, the dynamical behaviour of the solution can be accessed with a clever transfer-matrix trick, borrowed from the problem of Anderson localisation~\cite{Thouless1972Relation,Anderson1980New}. Let us first pass to the Fourier transform in time, ${\tilde{m}_x (\omega) = \int dt \, e^{-i \omega t} m_x(t)}$, which yields
\begin{multline}
    \label{eq:random_diffusion_Fourier}
    i\omega \tilde{m}_x (\omega) = D_{x-1} \tilde{m}_{x-1}(\omega) + D_x \tilde{m}_{x+1}(\omega) \\
    - (D_{x-1}+D_x) \tilde{m}_x(\omega).
\end{multline}
We now rewrite this equation in transfer matrix form:
\begin{equation}
    \begin{pmatrix}
        \tilde{m}_{x+1} \\
        \tilde{m}_x
    \end{pmatrix}
    =
    \begin{pmatrix}
        \frac{D_{x-1}+D_x+i\omega}{D_x}   &-\frac{D_{x-1}}{D_x} \\
        1                          &0
    \end{pmatrix}
    \begin{pmatrix}
        \tilde{m}_x \\
        \tilde{m}_{x-1}
    \end{pmatrix},
\end{equation}
which can be recast in a more compact notation,
\begin{equation}
    M_{x+1} = T_x(\omega) M_x,
\end{equation}
where we have introduced the vector of the two magnetisations $M_x$, and the $2 \times 2$ transfer matrix $T_x(\omega)$. Iterating, one finds
\begin{equation}
    \label{eq:iterated_transfer_matrix}
    M_{x+1} = T_x(\omega) T_{x-1}(\omega) \cdots T_1(\omega) M_1,
\end{equation}
which expresses $m_{x+1}$ for any $x$ in terms of $m_0$ and $m_1$.

Equation~\eqref{eq:iterated_transfer_matrix} does not admit a solution any more than the original form, but it does bring the problem into the realm of products of random matrices---a classic topic in statistical physics dating back to the works of Furstenberg~\cite{Furstenberg1960Products,Furstenberg1963Noncommuting}. Now, to get an idea of the nature of the product, let us consider the clean case,
\begin{equation}
  T=\begin{pmatrix}
        2+\frac{i\omega}{D}   &-1 \\
        1                          &0
    \end{pmatrix}.
\end{equation}
Even though $T$ is not Hermitian, it is diagonalisable and has eigenvalues
\begin{equation}
    \lambda_{1,2} = \frac{2D+i\omega \pm \sqrt{4iD\omega - \omega^2}}{2D}.
\end{equation}
Notice that, since $\det(T)=1$, it holds that $\lambda_{1} = 1/\lambda_2$; we choose the labels such that $|\lambda_1|\geq 1\geq|\lambda_2|$. Now, an $n$-fold application of $T$ to a generic vector corresponds (approximately) to a rotation plus an enlargement by a factor $|\lambda_1|$. This represents a vector localised {\it away} from the left boundary---ideally, on the right boundary. By, instead, fine-tuning the initial vector to the right eigenvector corresponding to $\lambda_2$, one finds a vector localised on the left boundary, which corresponds to the propagation of a disturbance created on the site $y=0$.   

The long-time dynamics corresponds to small values of $\omega$, for which 
\begin{equation}
    \lambda_{1,2} = 1 \pm \sqrt{\frac{i\omega}{D}} + \frac{i\omega}{2D} + \cdots .
\end{equation}
Therefore, for a disturbance localised on $y=0$, one finds
\begin{equation}
    \label{eq:left_localized}
    \tilde{m}_y(\omega) \simeq \left( 1 - \sqrt{i\omega/D} + \cdots \right)^y
    = e^{-y\sqrt{i\omega/D}},
\end{equation}
whilst the other eigenvalue corresponds to a disturbance localised at $y\to\infty$,
\begin{equation}
    \label{eq:right_localized}
    \tilde{m}_y(\omega) \simeq \left( 1 + \sqrt{i\omega/D} + \cdots \right)^y 
    = e^{+y\sqrt{i\omega/D}} .
\end{equation}
Above, we have set the values of the initial seed to $M_1 \approx 1$. The dispersion relation $y^2\sim Dt$ is immediately apparent from Eqs.~\eqref{eq:left_localized}--\eqref{eq:right_localized}, since $y$ and $\omega$ appear only in the combination $y\omega^{1/2}\sim y/t^{1/2}$. Then, selecting the decaying exponential, the inverse transform is
\begin{equation}
    m_y(t) \approx \int_{-\infty}^{+\infty} \frac{d\omega}{2\pi} e^{-\sqrt{i\omega/D} y + i\omega t}.
\end{equation}
If both $y$ and $t$ are large, this integral is dominated by the saddle point,
\begin{equation}
    0 = \pd{}{\omega} \left(-\sqrt{\frac{i \omega}{D}} y + i\omega t \right) \quad \implies \quad
    \omega = -i \frac{y^2}{4D t^2},
\end{equation}
and, substituting this back in, one finds
\begin{equation}
    m_y(t) \sim e^{-\frac{y^2}{4 D t}},
\end{equation}
i.e., diffusive behaviour $y^2 \approx 2Dt$. Note also that the gaussian tails of diffusion are correctly reproduced.

Let us now see how one can get the same results by expanding the product $T_x(\omega) T_{x-1}(\omega) \cdots T_1(\omega)$ in powers of $\omega$. This will be useful for the disordered case, as the eigenvalue of the product of the random transfer matrices cannot be obtained from the eigenvalues of the separate $T_x$'s. We consider the action of $T_x(\omega) T_{x-1}(\omega) \cdots T_1(\omega)$ on the trial vector $M_1=(1,1)^T$, which represents a good starting guess for a long-wavelength vector. Order-by-order in $\omega$, one finds (re-introducing the index on $D$ for future reference, although $D_x \equiv D$ in the clean case):
\begin{multline}
    \label{eq:m_expansion}
    \tilde{m}_{y+1}(\omega) = 1 + i\omega\sum_{x_1 \leq y}\frac{x_1}{D_{x_1}} 
    +(i\omega)^2 \sum_{x_1<x_2 \leq y} \frac{x_1 (x_2-x_1)}{D_{x_1} D_{x_2}} \\ 
    +(i\omega)^3 \sum_{x_1 < x_2 < x_3 \leq y} \frac{x_1(x_2-x_1)(x_3-x_2)}{D_{x_1} D_{x_2} D_{x_3}} + \cdots\ .
\end{multline}
The terms in the expression above are reminiscent of the locator expansion, which is a useful tool when studying the physics of localisation~\cite{Anderson1958Absence,AbouChacra1973Selfconsistent,Pietracaprina2016Forward}. 

The sums over $x_1,x_2,\dots$ in Eq.~\eqref{eq:m_expansion}, when the uniform $D$ is factored out, reduce to
\begin{equation}
    \sum_{x_1<\cdots < x_n \leq y} \!\!\!\!\!\!\!\! x_1 (x_2-x_1) \cdots (x_n - x_{n-1}) 
    = \frac{(y+n)!}{(2n)!(y-n)!} .
\end{equation}
The first terms in the large-$y$ expansion read
\begin{equation}
    \label{eq:Pochhammer_expansion}
    \frac{(y+n)!}{(2n)!(y-n)!} = \frac{y^{2n}}{(2n)!} \left( 1 + \frac{n}{y} + \cdots\right).
\end{equation}
Thus, in the limit of large $y$, one finds the approximate solution
\begin{align}
    \tilde{m}_y(\omega) &\approx \sum_{n=0}^\infty \frac{(i \omega)^n}{D^n} \frac{y^{2n}}{(2n)!}
    = \cosh \left( \sqrt{\frac{i \omega}{D}} y\right) \nonumber\\
    \label{eq:m_clean_cosh}
    &=\frac{1}{2}\left(e^{\sqrt{\frac{i \omega}{D}}y}+e^{-\sqrt{\frac{i \omega}{D}}y}\right),
\end{align}
the two terms corresponding exactly to the two eigenvalues of the transfer matrix. Again, selecting the decaying exponential and inverting the Fourier transform yields the diffusion profile and the Brownian dispersion relation.

To summarise, one can obtain the dispersion relation from the dependence of the series \eqref{eq:m_expansion} on the combination of $\omega$ and $y$, whilst access to the functional form requires the coefficients of the series. 

Crucially, this solution strategy can be transposed to the disordered case, as we now move to show in the following sections.

%--------------------------------------------------------------------------------------------------
\subsection{Scaling in the disordered model}
\label{sec:leading_scaling}

We now use the transfer-matrix method to solve the disordered chain, in which the $D_x$'s are i.i.d.\ random variables distributed according to $p_\eta(D)$. Note that, if the average $\overline{1/D}$ exists (i.e.\ $\eta < 0$), then $\overline{m_y}$ behaves diffusively, with an effective diffusion coefficient $D_\mathit{eff} = (\overline{1/D})^{-1}$. We stress that the effective diffusion coefficient is not given by $\overline{D}$, since the resistances $\sim 1/D_x$ are additive but the conductances $\sim D_x$ are not~\cite{Hulin1990Strongly}. 

The only case we need to treat, therefore, is when the moment $\overline{1/D}$ is infinite, i.e., $\eta \geq 0$. We will focus, however, on $\eta > 0$, leaving the limiting case $\eta = 0$ to \S\ref{sec:corrections_scaling}.

When $\eta > 0$, one must retain the explicit sums in Eq.~\eqref{eq:m_expansion}. Now, since $1/D_x$ has a fat-tailed distribution, the (finite) sums are dominated by the maximum---in particular,
\begin{equation}
    \label{eq:sum_maximum}
    \sum_{x_{i-1} \leq x_{i}} \frac{x_{i-1}}{D_{x_{i-1}}} \simeq \max_{x_{i-1}<x_{i}}\frac{x_{i-1}}{D_{x_{i-1}}}.
\end{equation}
For $\eta> 0$ the numerator is irrelevant---it is just a random number uniformly distributed in $[0,x_{i}]$, which we write as $c_i x_{i}$ with $c_i\in [0,1]$. Thus, simplifying further, one has
\begin{equation}
    \label{eq:sum_maximum}
    \sum_{x_{i-1} \leq x_{i}} \frac{x_{i-1}}{D_{x_{i-1}}} \simeq c_{i} x_{i}\max_{x_{i-1}<x_{i}}\frac{1}{D_{x_{i-1}}}.
\end{equation}
The maximum, over a large number of instances $x_i$, of the i.i.d.\ random variables $1/D_{x_{i-1}}$ is a random variable of typical value $x_{i}^{1/(1-\eta)} \gg x_{i}$, owing to $\eta>0$. Consequently, the whole sum is approximately
\begin{equation}
    \sum_{x_{i-1} \leq x_{i}} \frac{x_{i-1}}{D_{x_{i-1}}} \simeq b_{i} x_{i}^{1+\frac{1}{1-\eta}},
\end{equation}
where $b_i$ is another random variable of $O(1)$. Therefore:
\begin{gather}
    \sum_{x_{1} \leq y} \frac{x_1}{D_{x_{1}}} \simeq b^{(1)} y^{1+\frac{1}{1-\eta}}, \\
    \sum_{x_{1} <x_2\leq y} \frac{x_1(x_2-x_1)}{D_{x_{1}}D_{x_{2}}} \simeq b^{(2)} y^{2+2\frac{1}{1-\eta}},
\end{gather}
and so on.

We have now all the tools to evaluate the random series, Eq.~\eqref{eq:m_expansion}:
\begin{align}
    \tilde{m}_{y}(\omega)
    & = 1 + i\omega b^{(1)} y^{1+\frac{1}{1-\eta}} \big[1+o(y^0) \big] \nonumber\\ 
    &\ \phantom{=1} +(i\omega)^2 b^{(2)} y^{2+2\frac{1}{1-\eta}} \big[1+o(y^0) \big] \nonumber\\ 
    &\ \phantom{=1} +(i\omega)^3 b^{(3)} y^{3+3\frac{1}{1-\eta}} \big[1+o(y^0) \big] + \cdots\ \nonumber\\
    &= f_{\eta}(\omega y^{\frac{2-\eta}{1-\eta}}) \big[1+o(y^0) \big] .
    \label{eq:m_expansion_dis}
\end{align}
The neglected terms of $o(y^0)$ represent finite-time corrections, and they will be the object of the next section. The functional form of $f_\eta(x)$ cannot be evaluated at this coarse level of calculation, since it requires the knowledge of the coefficients $b^{(n)}$ at every order: for example, in the clean case one has $b^{(n)}=1/(2n)!$, and thus it simplifies to $f_\eta(x)=\cosh(\sqrt{i x})$ when $\eta\to-\infty$. 

Even if $f_\eta$ is left undetermined, the dispersion relation is found from the scaling
\begin{equation}
    t \sim \omega^{-1} \sim y^{(2-\eta)/(1-\eta)}
\end{equation}
or, equivalently, 
\begin{equation}
    \label{eq:scaling_eta>0}
    y \sim t^{(1-\eta)/(2-\eta)}.
\end{equation}
We conclude that, in the region $0<\eta<1$, the scaling is subdiffusive, with an exponent (Hurst index)
\begin{equation}
    \alpha = \frac{1-\eta}{2-\eta} < \frac{1}{2} .
\end{equation}
This is exactly the subdiffusive exponent used in \S\ref{sec:numerics_subdiffusion} (see Eq.~\eqref{eq:subdiffusion_autocorr} in particular) to fit the numerical data.

%--------------------------------------------------------------------------------------------------
\subsection{Finite-time corrections to the scaling}
\label{sec:corrections_scaling}

As can be seen from the numerical data of \S\ref{sec:numerics}, sizeable corrections to the asymptotic scaling persist until very long times in the bond-disordered Heisenberg chain, Eq.~\eqref{eq:Heisenberg_Hamiltonian}. This feature is shared by the phenomenological model, Eq.~\eqref{eq:random_diffusion}, as we now show. We will split the discussion for the regimes of diffusion, slow diffusion, and subdiffusion; the logarithmically-suppressed diffusion will follow as a limiting case.

\paragraph{Diffusion.} To set the stage, let us first address the finite-time corrections in the clean case $D_x \equiv D$ (i.e.\ $\eta \to -\infty$). The same features are shared by the whole region $-\infty < \eta < -1$, as will become clear. Retaining the first-order corrections in Eq.~\eqref{eq:Pochhammer_expansion}, one can resum the series in Eq.~\eqref{eq:m_expansion} to
\begin{equation}
    \tilde{m}_{y+1}(\omega) = \cosh \left( \sqrt{\frac{i \omega}{D}} y\right) + \frac{1}{2} \sqrt{\frac{i \omega}{D}} \sinh \left( \sqrt{\frac{i \omega}{D}} y\right) + \cdots
\end{equation}
The second term, upon taking the inverse Fourier transform, is dominated by the same saddle as the first, and one finds
\begin{equation}
    m_{y+1}(t) = e^{-\frac{y^2}{4 D t}} \left( \frac{1}{2} + \frac{y}{8Dt} + \cdots \right).
\end{equation}
Again, the overall constant needs to be fixed by normalisation, since the initial guess for $m$ was not normalised. What counts for our purposes is the relative size of the first two terms: using $y\sim t^{1/2}$ from the scaling, the second term is seen to be of order $t^{-1/2}$ w.r.t.\ the first.

Upon reintroducing the disorder, finding the explicit first-order corrections to Eq.~\eqref{eq:m_expansion_dis} is more difficult, and a careful study of the random sums in Eq.~\eqref{eq:m_expansion} at all orders of $\omega$ is needed. Indeed, $\tilde{m}_y(\omega)$ is itself a random variable, and the large-space and long-time behaviour of $m_y(t)$ should be inspected by considering not only the average, $\overline{m_y(t)}$, but also its moments---or, equivalently, the average of quantities such as $\overline{\log m_y(t)}$. For this reason, we find it convenient to pass to the logarithm \emph{at the level of the Fourier transform}:
\begin{multline}
    \label{eq:m_expansion_log}
    \log \tilde{m}_{y+1}(\omega) = i\omega\sum_{x_1 \leq y}\frac{x_1}{D_{x_1}} \\
    +(i\omega)^2 \bigg[ \sum_{x_1 \leq y} \frac{x_1^2}{D_{x_1}^2}
    + 2 \sum_{x_1 < x_2\leq y} \frac{x_1^2}{D_{x_1} D_{x_2} } \bigg] + \cdots\ 
\end{multline}
The equation above has the useful property that, at each order $\omega^n$, there is one term $\propto 1/D_{x_1}^n$, followed by less singular terms $1/D_{x_1}^{n-1}D_{x_2}$, $1/D_{x_1}^{n-2}D_{x_2} D_{x_3}$, and so on. When the moment $\overline{1/D^n}$ does not exist, but all the moments $\overline{1/D^m}$ with $m<n$ exist (i.e.\ for $\eta \leq -n+1$), an anomalous contribution to $\tilde{m}_y(\omega)$ appears---influencing the finite-time dynamics at order $\omega^n$. As long as $\eta < -1$, both the terms of order $\omega$ and $\omega^2$ are regular, and thus we expect moderately long times to suffice to make diffusion manifest. On the other hand, when $\eta$ crosses $-1$, the first correction $O(\omega^2)$ gains an anomalous power, and signatures of slow diffusion are found. We detail this fact in the next paragraph.

\paragraph{Slow diffusion.} Let us focus again on Eq.~\eqref{eq:m_expansion_log}. By using the same analysis as \S\ref{sec:leading_scaling} for the sums of random variables, one finds, in the region $-1 \leq \eta < 0$,
\begin{gather}
    \label{eq:sum_x1}
    \sum_{x_1 \leq y}\frac{x_1}{D_{x_1}} \sim \overline{1/D} \, y^2,\\
    \label{eq:sum_x1^2}
    \sum_{x_1 \leq y} \frac{x_1^2}{D_{x_1}^2} \sim y^{2(2-\eta)/(1-\eta)}, \\
    \label{eq:sum_x1x2}
    \sum_{x_1 < x_2\leq y} \frac{x_1^2}{D_{x_1} D_{x_2} } \sim \overline{1/D}^2 y^4,
\end{gather}
and similarly for higher moments. By looking at the expression above, one recognizes that the terms~\eqref{eq:sum_x1} and \eqref{eq:sum_x1x2} combine to form a regular function of $\omega y^2$. Indeed, similarly to the subdiffusive case (see Eq.~\eqref{eq:m_expansion_dis}), one can group terms and find
\begin{equation}
    \label{eq:logm_slow_diff}
    \log \tilde{m}_{y+1}(\omega) = \log f_{\eta}(\omega y^2) + c^{(1)}\omega^2 y^{2(2-\eta)/(1-\eta)}+ \cdots
\end{equation}
with $c^{(1)}$ being a constant of $O(1)$. This has to be interpreted in the same way as an anomalous scaling of the free energy at a critical point: the analytic part is represented by the first term and it is followed by a series of anomalous corrections, beginning with $\omega^2 y^{2(2-\eta)/(1-\eta)}\sim t^{\eta/(1-\eta)}$ (having used the scaling of the dominant term $y^2\sim t$).

We remark that the subleading terms are very important when one wants to extract the scaling exponents from the numerics, as we already showed in \S\ref{sec:numerics}. If they are not properly accounted for, the errors are rather large and the determination of the onset of subdiffusion is misplaced. The reason is that the subleading term $t^{\eta/(1-\eta)}$ has to be much smaller than unity if one wants to extract the leading exponent $1/2$ with some accuracy: this requires extremely long times for $|\eta| < 1$, and is a major source of obfuscation in the analysis of numerical data as shown in the previous sections. 

\paragraph{Subdiffusion.} We now consider the case $\eta > 0$, where not even the first moment of the random variable $1/D$ exists. Extreme-value statistics tells us that all the sums in Eqs.~\eqref{eq:m_expansion} or \eqref{eq:m_expansion_log} become anomalous, giving rise to the expression in Eq.~\eqref{eq:m_expansion_dis}. Here, we argue that the corrections left out in Eq.~\eqref{eq:m_expansion_dis} involve \emph{regular} powers of $y$, as we illustrate with a very simple example. Consider the first-order term $i \omega \sum_{x_1 \leq y} x_1 / D_{x_1}$. Let us split the sum according to whether $D_{x_1} > D_\star$ or $D_{x_1} < D_\star$, where the value $D_\star$ is fixed so that the probability $D<D_\star$ is $p=1/2$ (any other finite value of $p$ would lead to the same conclusion). Then, one recognises that the random variable
\begin{equation}
    \psi:=\sum_{x_1 \leq y} \frac{x_1}{D_{x_1}} 
\end{equation}
has a broad probability distribution peaked at a value $\psi \sim y^{(2-\eta)/(1-\eta)}$, but with non-zero weight down to $\psi\sim y^2$: this latter value comes from the regular sum of the terms involving $D_{x_1} > D_\star$, whilst the former represents the anomalous contribution of the very small instances $D_{x_1} < D_\star$. So, with a slight abuse of notation, one can say that 
\begin{equation}
    \sum_{x_1 \leq y} \frac{x_1}{D_{x_1}}\sim y^{(2-\eta)/(1-\eta)}+c y^2,
\end{equation}
in the sense that all functions of this random variable may be expanded, at large $y$, in these two (leading and subleading) terms.

A careful treatment of the random sums thus leads to two families of terms: those involving regular powers of $y$, and those involving anomalous powers. These two families receive contributions from all orders in $\omega$, and a resummation of all the terms is beyond the scope of this work. We will content ourselves with the following simple scaling analysis: the combination of $\omega$ and $y$ which appears is
\begin{equation}
    \omega (y^{(2-\eta)/(1-\eta)} + c y^2) \sim 1,
\end{equation}
from which it follows that 
\begin{equation}
    y \sim t^\alpha (1 + c' t^{2\alpha - 1}),
    \label{eq:logm_subdiffusion}
\end{equation}
which is the form of the autocorrelator \eqref{eq:subdiffusion_autocorr} used to fit the numerical data. The constants $c$ and $c'$ cannot be fixed at this rough level of calculation; thus, in \S\ref{sec:numerics}, some fitting was still required.

\paragraph{Logarithmically-suppressed diffusion.} We finally consider how logarithmically-suppressed diffusion emerges. Being the limiting case between subdiffusion and slow diffusion, it can be understood from both sides. From the slow-diffusion side, one can see that the corrections to the asymptotic (diffusive) scaling tend to become of the same order of the leading term as $\eta \to 0^-$: this is because $(2-\eta)/(1-\eta) \to 2$, and the two terms on the r.h.s.\ of Eq.~\eqref{eq:logm_slow_diff} coalesce, forming a logarithm. The same happens from the subdiffusive side, where the dominant term is now $y^{(2-\eta)/(1-\eta)}$, whilst the corrections are given by $y^2$: the mechanism is the same, though the role of the two terms is exchanged. We point out that this coalescence of power-laws can be understood from a complementary perspective via the integral-equation solution, see App.~\ref{app:sec:integral_eq}.

%--------------------------------------------------------------------------------------------------
%--------------------------------------------------------------------------------------------------
\section{Conclusions}
\label{sec:conclusions}

We have shown that the dynamics of a classical Heisenberg chain with broadly-distributed couplings $J_i$, specifically $p_{\eta}(J)\sim J^{-\eta}$, goes through various dynamical phases as $\eta$ is increased from very negative to its maximum achievable value, $\eta=1$. For $\eta<-1$ the correlation functions are diffusive (data not presented, though see, e.g., the supplementary material of Ref.~\cite{McRoberts2022Anomalous}). For $-1<\eta<0$, we have shown that, whilst the asymptotic behaviour is still diffusive, there are strong finite-time corrections which can be mistaken as signs of subdiffusive transport. True subdiffusion sets in only when $\eta>0$, with the subdiffusive exponent matching the analytic prediction of a phenomenological model in which a local diffusion coefficient is assumed to be a random variable, also broadly-distributed, with the same exponent $\eta$ as the local couplings $J$.

We point out that the quantum version of the model considered here \eqref{eq:Heisenberg_Hamiltonian} was the subject of recent works~\cite{Protopopov2020NonAbelian,Saraidaris2023Finite}, in which it was argued that a regime intermediate between many-body localisation and thermalisation persists in the thermodynamic limit. Such a regime is found in a range of parameters equivalent to our $0 \leq \eta <1$, i.e., when the classical model shows subdiffusive transport. It may be interesting to consider whether a semiclassical treatment of the quantum model could link these findings.

Our work, we believe, settles the question about the onset of diffusion and subdiffusion in classical Heisenberg chains with random couplings, in part already considered in Refs.~\cite{Srivastava1994Spin,Bagchi2013Spin}. It also presents yet another cautionary tale for efforts to extract potentially anomalous dynamical exponents, and identify possible dynamical phase transitions based on short-time, small-system numerics. Indeed, the discrepancy between exponents obtained from different but, visually, similarly good fits on system sizes of several thousand spins at times of a million $J_{\max}^{-1}$, gives a quantitative indication of just how challenging it is to estimate ``systematic'' error bars.

To conclude, leaving aside the considerations of a largely technical nature, the family of models we have studied provides a window on the physics of how rare (or not-so-rare) local fluctuations manifest themselves at long length- and time-scales. Our work, in this sense, is a classical counterpart to the strong-disorder renormalisation group treatments~\cite{Ma1979Random,Dasgupta1980Low,Fisher1995Critical,Refael2013Strong,Igloi2018Strong} which have been so influential for the study of quantum models in the last few decades. The question of which regimes still await discovery, in addition to those found and referenced in this work, strikes us as a subject of study likely to hold more than one surprise in store. 

%--------------------------------------------------------------------------------------------------
%--------------------------------------------------------------------------------------------------
\acknowledgements

F.B.\ would like to thank Federica Ferretti, Alessandro Manacorda and Carlo Vanoni for discussion. A.J.McR.\ and F.B.\ thank ICTP for hospitality during the completion of this work. A.S.\ acknowledges financial support from PNRR MUR project PE0000023-NQSTI. This work was in part supported by the Deutsche Forschungsgemeinschaft under grant  cluster of excellence ct.qmat (EXC 2147, project-id 390858490). 

%--------------------------------------------------------------------------------------------------
%--------------------------------------------------------------------------------------------------
\appendix

%--------------------------------------------------------------------------------------------------
%--------------------------------------------------------------------------------------------------
\section{Integral equation solution of the effective model}
\label{app:sec:integral_eq}

In this appendix we solve the effective model, Eq.~\eqref{eq:random_diffusion}, taking inspiration from the calculations of Ref.~\cite{Bernasconi1980Diffusion}, but employing a simpler strategy. We start by rewriting Eq.~\eqref{eq:random_diffusion} as
\begin{equation}
    \partial_t m_x(t) = - (H m)_x(t),
\end{equation}
where 
\begin{equation}
    H = \sum_x \big[ (D_x + D_{x-1}) \ket{x}\bra{x} - D_x ( \ket{x+1}\bra{x} + \mathrm{h.c.}) \big],
\end{equation}
and the state $\ket{x}$ represents a particle located at position $x$ on the chain. We now introduce the diagonal element of the resolvent of $H$, namely,
\begin{equation}
    \label{app:eq:def_resolvent}
    G_{00} := \bra{0} \frac{1}{\omega-H} \ket{0}.
\end{equation}
Notice that here the frequency $\omega$ is obtained via a Laplace transform instead of a Fourier transform, so there is an imaginary unit of difference w.r.t.\ \S\ref{sec:effective_model}.

\begin{figure}[t]
    \centering
    \includegraphics[width=\columnwidth]{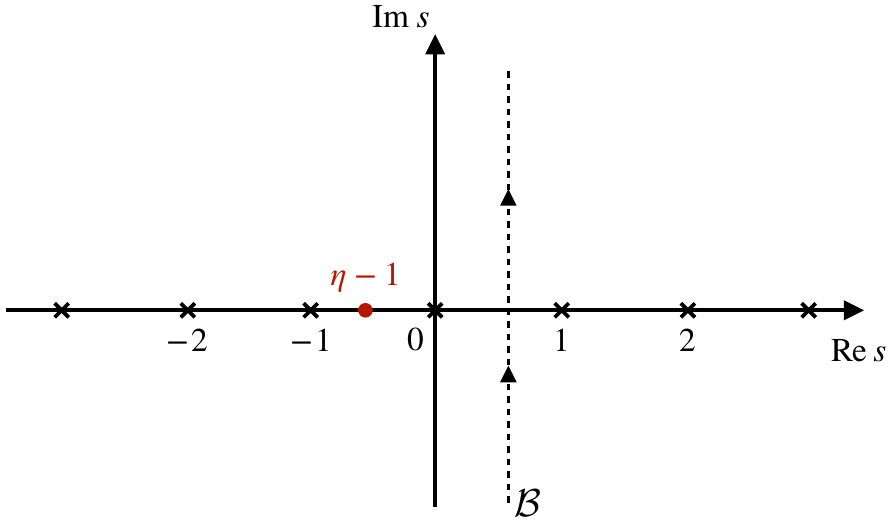}
    \caption{Bromwich contour for the inversion of the Mellin transform, Eq.~\eqref{eq:F_Mellin}. The poles at integer-values (black crosses) are responsible for a regular scaling of time and space, whilst the pole at $s=\eta-1$ (red dot) is responsible for subdiffusion when $\eta >0$, and for the anomalous corrections to diffusion when $-1 < \eta < 0$.}
    \label{fig:Bromwich}
\end{figure}

Using standard methods for tridiagonal matrices---or, equivalently, a locator expansion~\cite{Anderson1958Absence,AbouChacra1973Selfconsistent,Pietracaprina2016Forward}---Eq.~\eqref{app:eq:def_resolvent} can be recast in the form
\begin{equation}
\label{app:eq:G00}
    G_{00}(\omega)=\frac{1}{G_+ +G_-+\omega},
\end{equation}
with the random variables $D_x$ to the right of site 0 appearing in 
\begin{equation}
    G_+=\frac{1}{D_1^{-1}+\frac{1}{\omega+\frac{1}{D_{2}+...}}},
\end{equation}
and those to the left appearing in
\begin{equation}
    G_-=\frac{1}{D_0^{-1}+\frac{1}{\omega+\frac{1}{D_{-1}+...}}}.
\end{equation}
Now, $G_{\pm}$ are themselves random variables, and their distribution can be found in an iterative way. In fact, the relation
\begin{equation}
    G_{\pm,x}=\frac{1}{D_x^{-1}+1/(\omega+G_{\pm, x-1})}
\end{equation}
is a sort of recursion equation familiar from the theory of Anderson localisation \cite{AbouChacra1973Selfconsistent,Pietracaprina2016Forward,Parisi2019Anderson}, and that of spin glasses \cite{Mezard2001Bethe,Laumann2008Cavity,Laumann2008Griffiths}. The limiting distribution of $G$ must be invariant under the iteration
\begin{multline}
    f(g)=\int dg'f(g')\int dD \, \rho(D) \\
    \times \delta\left[ g-\left(D^{-1}+(\omega+g)^{-1}\right)^{-1} \right].
\end{multline}
The scaling form at small $\omega$ can be recovered by looking at the first moment $\overline{g} = \int dg \,g \, f(g)$:
\begin{align}
\label{app:eq:mean_g}
    \overline{g} &=\int dg' f(g') \int dD \, \rho(D) \left[D^{-1}+(\omega+g)^{-1}\right]^{-1} \nonumber\\
    &=\int dg' f(g')(\omega+g')F(\omega+g'),
\end{align}
where
\begin{align}
    F(s)&=\int dD \,\rho(D)\frac{1}{1+s/D}\nonumber\\
    &=(1-\eta)\int_1^\infty dR \, R^{-2+\eta}\frac{1}{1+sR},
\end{align}
having passed to the variable $R:=1/D$. Notice that $F(0)=1$, but one needs also the corrections at small $s=\omega+g'$. Going to the Mellin transform, one can write
\begin{equation}
    \label{eq:F_Mellin}
    F(s)=\int_\mathcal{B}\frac{dz}{2\pi i}\frac{\pi s^{-z}}{\sin{\pi z}}\frac{1-\eta}{z+1-\eta},
\end{equation}
\begin{figure}
    \centering
    \includegraphics{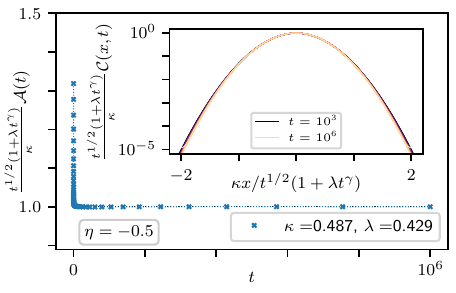}
    \caption{Slow diffusion in the effective model at $\eta = -0.5$, obtained by directly simulating Eq.~\eqref{eq:random_diffusion}. The leading and sub-leading terms predicted by the transfer matrix solution, and used to fit the spin correlations in the main text for the same value of $\eta$ (Fig.~\ref{fig:slow_diffusion}), provide an excellent fit to the data.}
    \label{fig:app_slow_diffusion}
\end{figure}
where $\mathcal{B}$ is the Bromwich path from $-i\infty$ to $+i\infty$ with $0<\Re(z)<1$, see Fig.~\ref{fig:Bromwich}. The function $F(s)$ in the complex $s$ plane contains poles at all the integers $s\in \mathbb{Z}$ and at $s=\eta-1<0$, see again Fig.~\ref{fig:Bromwich}. In order to find the small-$s$ behaviour, one can move the contour to the left, picking up as many poles as terms required. For $\eta>0$, one finds
\begin{equation}
    \label{app:eq:F_poles}
    F(s)=1-\frac{\pi(1-\eta)}{\sin{\pi(1-\eta)}}s^{1-\eta}+O(s).
\end{equation}
Inserting the relation above in Eq.~\eqref{app:eq:mean_g}, we have
\begin{align}
    \overline{g} &= \int dg' f(g')(\omega+g')\left[1-\frac{\pi(1-\eta)}{\sin{\pi(1-\eta)}}(\omega+g')^{1-\eta}\right] \nonumber\\
    &=\omega+ \overline{g} - \frac{\pi(1-\eta)}{\sin{\pi(1-\eta)}} \overline{(\omega+g)^{2-\eta}},
\end{align}
with the promised small $s$ corrections. Neglecting, self-consistently, $\omega$ w.r.t.\ $g$, one obtains
\begin{equation}
    \omega = \frac{\pi(1-\eta)}{\sin{\pi(1-\eta)}} \overline{g^{2-\eta}},
\end{equation}
and so
\begin{equation}
    \overline{g^{2-\eta}} \sim \omega.
\end{equation}
Analogously, for all $n\geq 2$ one can prove that $\overline{g^{n-\eta}} / \overline{g^{n-2}} \sim \omega$. Therefore, the typical value of $g \sim \omega^{1/(2-\eta)}$ which, when inserted in Eq.~\eqref{app:eq:G00}, gives
\begin{equation}
    \mathcal{A}(t)\sim\int d\omega\frac{e^{i\omega t}}{\omega^{\frac{1}{2-\eta}}+O(\omega)}\sim t^{-\frac{1-\eta}{2-\eta}}.
\end{equation}
This is consistent with the result obtained in the main text.

From the Mellin transform formalism one can also get a complementary understanding of the subleading terms. Looking at Eq.~\eqref{eq:F_Mellin} or Fig.~\ref{fig:Bromwich}, one can see that the function $F(s)$ receives contributions from two kinds of poles: those at integer-values, and an anomalous pole at $s = \eta -1$. This last pole moves as the disorder strength $\eta$ is tuned, and, depending on the relative position of the anomalous pole $s =\eta -1$ and the pole at $s = -1$, the asymptotic behaviour changes from diffusion to subdiffusion: indeed, it is the first pole to the left of $s=0$ that determines the asymptotics. One can see that the two poles coalesce precisely at $\eta = 0$, in accordance with the power series treatment of \S\ref{sec:effective_model}.

The poles at $s=-2$, $s=-3$, etc., represent subleading corrections to the scaling of $\overline{g}$ w.r.t.\ $\omega$. When $-1 < \eta < 0$, i.e., in the slow diffusion regime, all such poles are subleading; the anomalous pole is the first to be encountered to the left of $s=-1$, and thus provides the leading finite-time corrections to the asymptotic scaling. When, instead, $\eta < -1$, it is the pole at $s=-2$ that dominates the corrections, and standard diffusion is recovered to a very good approximation.

%--------------------------------------------------------------------------------------------------
%--------------------------------------------------------------------------------------------------
\section{Numerics of the effective model}
\label{app:sec:effective_model}

\begin{figure}
    \centering
    \includegraphics{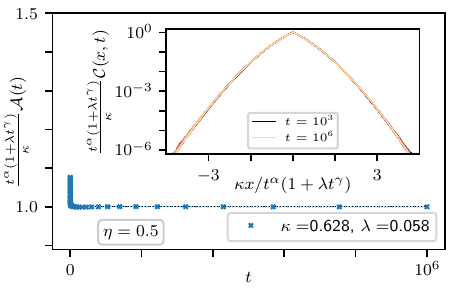}
    \caption{Subdiffusion in the effective model at $\eta = 0.5$, obtained by directly simulating Eq.~\eqref{eq:random_diffusion}. Similarly to the case of slow-diffusion (Fig.~\ref{fig:app_slow_diffusion}), the transfer-matrix solution, which describes the spin correlations at the same $\eta$ (cf. Fig.~\ref{fig:subdiffusion}), is in agreement with the effective model's numerics.}
    \label{fig:app_subdiffusion}
\end{figure}

As a further consistency check that our effective model accurately captures the phenomenology of the disordered Heisenberg chains \eqref{eq:Heisenberg_Hamiltonian}, and that our transfer-matrix approach is correct, we directly simulate Eq.~\eqref{eq:random_diffusion}. We take a finite system with sites $x \in [-L/2, L/2)\cap\mathbb{Z}$, $L = 8192$, and the initial conditions $m_{x} = \delta_{x, 0}$. The results are averaged over $20 000$ realisations of the random coefficients $D_x$. We expect the spreading of the magnetisation profile $m_x(t)$ under this set-up to mimic the spin correlations of Eq.~\eqref{eq:Heisenberg_Hamiltonian}.

\begin{figure}
    \centering
    \includegraphics{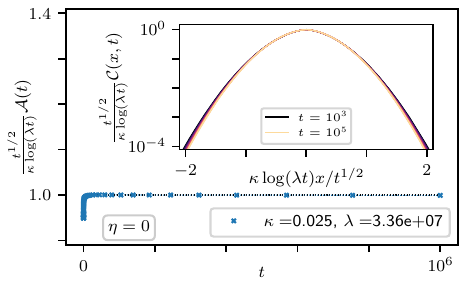}
    \caption{Logarithmically suppressed diffusion at $\eta = 0$ in the effective model Eq.~\eqref{eq:random_diffusion}. }
    \label{fig:app_log_diffusion}
\end{figure}

We show convincingly in Fig.~\ref{fig:app_slow_diffusion} that the slow-diffusion regime ($-1 < \eta < 0$) falls within these expectations, with the leading and sub-leading terms from the transfer matrix solution \eqref{eq:logm_slow_diff} providing an excellent fit to the autocorrelator, and a scaling collapse over three decades of time. We find that the subdiffusive regime $\eta > 0$ evinces a similarly good agreement, shown in Fig.~\ref{fig:app_subdiffusion}, using the leading and sub-leading terms found in Eq.~\eqref{eq:logm_subdiffusion}. Finally, the transition point $\eta = 0$, as displayed in Fig.~\ref{fig:app_log_diffusion}, shows a good agreement with the autocorrelator, and a decent scaling collapse. We note that the size of the parameter $\lambda \approx 10^7$ does not imply a poorly converged fit, since the combination $\kappa \log \lambda \approx 0.433$ entails a perfectly-reasonable, additive, subleading correction.

%--------------------------------------------------------------------------------------------------
\bibliography{references}

\end{document}